\documentclass[twocolumn,secnumarabic, showpacs, superscriptaddress,amssymb,english,aps, prb,floatfix, longbibliography]{revtex4-1}

\usepackage{graphicx}
\usepackage{SIunits}
\usepackage{epstopdf}

\begin{document}

\title{Temperature-dependent photoluminescence characteristics of GeSn epitaxial layers}

\author{Fabio Pezzoli}
\email[]{fabio.pezzoli@unimib.it}
\affiliation{LNESS and Dipartimento di Scienza dei Materiali, Universit\`a  degli Studi di Milano-Bicocca, via Cozzi 55, I-20125 Milano, Italy}

\author{Anna Giorgioni}
\affiliation{LNESS and Dipartimento di Scienza dei Materiali, Universit\`a  degli Studi di Milano-Bicocca, via Cozzi 55, I-20125 Milano, Italy}

\author{David Patchett}
\affiliation{Department of Physics, The University of Warwick, Coventry CV4 7AL, United Kingdom}

\author{Maksym Myronov}
\affiliation{Department of Physics, The University of Warwick, Coventry CV4 7AL, United Kingdom}

\begin{abstract}
$\mathrm{Ge_{1-x}Sn_x}$ epitaxial heterostructures are emerging as prominent candidates for the monolithic integration of light sources on Si substrates. Here we propose a judicious explanation for their temperature-dependent photoluminescence (PL) that is based upon the so far disregarded optical activity of dislocations. By working at the onset of plastic relaxation, which occurs whenever the epilayer releases the strain accumulated during growth on the lattice mismatched substrate, we demonstrate that dislocation nucleation can be explicitly seen in the PL data. Notably, our findings point out that a monotonous thermal PL quenching can be observed in coherent films, in spite of the indirect nature of the $\mathrm{Ge_{1-x}Sn_x}$ bandgap. Our investigation, therefore, contributes to a deeper understanding of the recombination dynamics in this intriguing group IV alloy and offers insights into crucial phenomena shaping the light emission efficiency. 
\end{abstract}

\maketitle

The recently discovered lasing action in $\mathrm{Ge_{1-x}Sn_x}$ \cite{Wirths15}, tensile-strained \cite{Liu10, Camacho12} and nanostructured \cite{Grydlik16} Ge has sparked the quest for identifying group IV candidates boasting a direct bandgap. Their monolithic integration on Si holds the promise to revolutionize information technology by spurring the wafer-scale establishment of advanced photonic circuits.\cite{Liang10} Arguably, light-routing into the cost-effective Si platform has applications beyond communication, including biological sensing, \cite{Chow04, Soref10} optical memories \cite{Ferrera10, Rios15} and quantum computation.\cite{Politi08, OBrien09, Shadbolt14}

Despite the remarkable progress in heteroepitaxy and fabrication techniques that made strain and bandgap engineering readily available, the contrasting theoretical\cite{Fitzgerald91, He97, Moontragoon07} and experimental \cite{Ryu13, Ghetmiri14, Gallagher14, Wirths15} data accumulated in the literature proved the crucial resolution of the indirect-to-direct crossover to be non-trivial and highly debated in semiconductors based on group IV materials. 

The increase of the photoluminescence (PL) intensity with decreasing temperature has been recently put forward as a compelling proof that the fundamental energy gap of $\mathrm{Ge_{1-x}Sn_x}$ binary alloys can be direct in the momentum space.\cite{Wirths15} Since this method is now increasingly applied, \cite{Stange15, Virgilio15, ElKurdi16, Biswas16, Geiger16} we find it crucial to examine the possible pitfalls associated with it. 
Indeed, the PL intensity is, to a good approximation, proportional to the quantum efficiency, thus prone to the entangled interplay between all the possible radiative and nonradiative processes. While the PL enhancement with decreasing temperature can be intuitively associated to direct bandgap semiconductors, which entail III-V compounds and transition metal dichalcogenide monolayers as well-cited examples, under suitable conditions it can unexpectedly occur also in indirect bandgap materials\cite{Klein12} including Ge.\cite{Wan01, Kamenev06, Lee08, Pezzoli14}  

In this work, we demonstrate that a detailed knowledge of the recombination dynamics at ubiquitous defects, namely dislocations, is pivotal for a correct physical interpretation of the measured temperature dependence of the PL. Notably, despite the evidences that are arising for defect-related Shockley-Read-Hall recombination in group IV heterostructures,\cite{Wirths15, Gallagher15} yet very little is known about the nature and the optical activity of defects, chiefly dislocations. Understanding such phenomena is indeed of prime interest for overcoming the stringent roadblocks in the achievement of efficient lasing operation at room-temperature. 

To tackle this challenge we focus on ultrathin $\mathrm{Ge_{1-x}Sn_x}$ epitaxial layers grown on Ge buffered Si substrates. By deliberately exploiting alloys with a Sn content below 9\%, \cite{Wirths15} we restrict ourselves to indirect gap materials and leverage their optical response  as very sensitive probes of the defect-assisted recombination and generation of the photo-generated carriers. Notably, varying the epilayer thickness at a fixed Sn content provides us with samples with an adjustable strain relaxation. This eventually paves the way to  suppress the plastic relaxation, selectively inhibiting dislocation nucleation. By doing so, we are well-positioned to explore how dislocations affect recombination dynamics through the opening of parasitic nonradiative recombination channels. 

\begin{table*}\caption{List of the $\mathrm{Ge_{1-x}Sn_x}$ samples investigated in this work.}
  \vspace{0.1cm}
  \label{tab:samples}
  \tabcolsep=0.49cm
  \renewcommand{\arraystretch}{1.5}
  \begin{tabular}{c|cc}
    \hline \hline
    Sn molar fraction (\%) &  $\mathrm{Ge_{1-x}Sn_x}$ thickness  (nm) & Strain relaxation degree (\%) \\ \hline \hline
    5 & 70 & 0 \\
    5 & 100 & 11.5 \\
    9 & 40 & 0 \\
		9 & 70 & 0.6 \\
    9 & 80 & 6 \\ \hline \hline
  \end{tabular}
\end{table*}

\begin{figure}
\includegraphics[width=8cm]{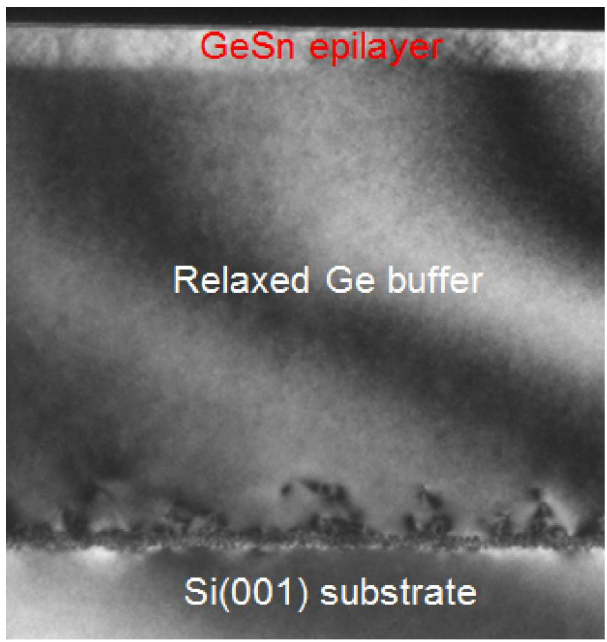}
  \vspace{0.5cm}
 \caption{Cross-sectional TEM image of a fully strained $\mathrm{GeSn}$ epilayer. The silicon substrate, and the Ge buffer layer are also indicated.}
	\label{fig1}
\end{figure}


For this work the $\mathrm{Ge_{1-x}Sn_x}$ epilayers of various Ge content and thickness were grown on an 100 mm diameter Si (001) substrate via an intermediate relaxed Ge buffer layer to minimise the lattice mismatch, and consequently the compressive strain, between the $\mathrm{Ge_{1-x}Sn_x}$ epilayer and the Si substrate. The structures were grown using industrial type ASM Epsilon 2000 reduced pressure chemical vapour deposition (RP-CVD) reactor with commercially available precursors digermane ($\mathrm{Ge_2H_6}$) and tin-tetrachloride ($\mathrm{SnCl_4}$) diluted in Hydrogen.  The epitaxial growth was carried out in Hydrogen atmosphere and at reduced pressure below 100 Torr. The Sn fractions in the $\mathrm{Ge_{1-x}Sn_x}$ epilayers and their state of strain were obtained from analysis of symmetrical (004) and asymmetrical (224) HR-XRD reciprocal space maps. The epilayers thicknesses were obtained by analysis of cross-sectional Transmission Electron Microscopy (XTEM) images. Representative XTEM data are shown in Figure 1, while the complete survey of the sample parameters is given in Table 1. It should be noted that just $\sim 650$ nm relaxed Ge buffer layer was grown on Si substrate to produce very high quality surface with RMS surface roughness below 1 nm and TDD density below $10^7$ $\mathrm{cm^{-2}}$ for subsequent growth of $\mathrm{Ge_{1-x}Sn_x}$ epilayers.  The compressive-strained (or coherent) $\mathrm{Ge_{1-x}Sn_x}$ epilayers are free from defects apart from threading dislocations arising from the Ge buffer layer.  However, XTEM images of partially relaxed epilayers shows appearance of Lomer $\mathrm{90^o}$-dislocations at the $\mathrm{Ge_{1-x}Sn_x}$/Ge interface. 


To gain insight into the optical properties of $\mathrm{Ge_{1-x}Sn_x}$ epitaxial layers, PL measurements were carried out using a closed-cycle variable temperature cryostat. The samples were excited at 1.165 eV (1064 nm) by using a Nd-YVO$_4$ laser. The optically pumped surface area had approximately a circular shape with a diameter of $\sim 40\mathrm{\mu m}$, which resulted in a power density of few $\mathrm{kW/cm^2}$. The PL was analyzed by a monochromator equipped with an InGaAs detector and double-checked by a Fourier transform spectrometer coupled to a PbS detector. In the former case, the PL spectra were numerically cleaned to remove the overlap with the second order peak due to the incomplete rejection of the pump.

We begin by focusing on the optical properties of the samples with the highest Sn-content. Figure 2a demonstrates the temperature-dependent PL of the sample capped with the partially relaxed $\mathrm{Ge_{0.91}Sn_{0.09}}$ film. In line with recent literature works, the peak at $\sim 0.6$ eV is attributed to the interband optical transitions occurring in the topmost $\mathrm{Ge_{1-x}Sn_x}$ alloy layer.  Notably, the PL signal turns out to be sizable up to room temperature, despite the overall reduced thickness of the $\mathrm{Ge_{1-x}Sn_x}$/Ge stack and the complete absence of any surface passivation. At low temperature, the PL intensity increases by decreasing the relaxation degree (Figure 2a-b). Eventually, the thinnest and fully-strained $\mathrm{Ge_{0.91}Sn_{0.09}}$ counterpart demonstrates the brightest emission (Figure 2c): A result which already discloses that the attained coherent growth suppresses nonradiative bulk-like recombination pathways, whose role, in addition, can be inferred to be prominent over the competing channel offered by the surface states. 

\begin{figure}
\includegraphics[width=8.5cm]{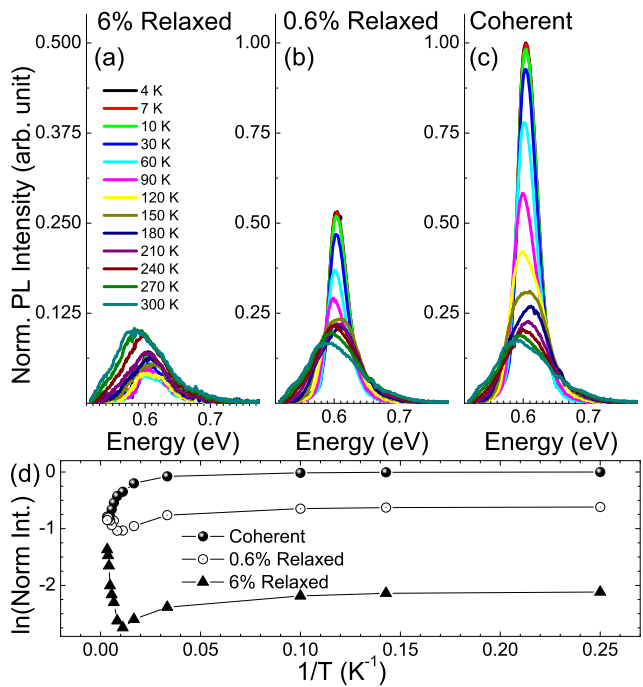}
 \caption{Photoluminescence (PL) spectra as a function of the temperature for $\mathrm{Ge_{0.91}Sn_{0.09}}$ layers epitaxially grown on Ge buffered Si(001) substrates. According to X-ray diffraction measurements, the epilayers are either partially relaxed with a relaxation degree of 6\% (a) and 0.6\% (b) or (c) fully strained, being grown coherently with the substrate. (d) Arrhenius plot summarizing the temperature dependence of the integrated PL intensity for the 6\% (triangles), 0.6\% (open dots) partially relaxed and coherent (full dots) films. All the data are normalized to the low temperature emission intensity of the coherent $\mathrm{Ge_{0.91}Sn_{0.09}}$ sample.}
	\label{fig2}
 \end{figure}

It should be noted that no PL from the underlying Ge buffer was observed at the typical emission energies of Ge-on-Si heterostructures in the range from 0.6 to 0.9 eV. \cite{Pezzoli14} The vast majority of radiative recombination events occurring in the $\mathrm{Ge_{1-x}Sn_x}$ layer is thus likely to be due to its strong light absorption and the favorable $\mathrm{Ge_{1-x}Sn_x}$/Ge band edge alignment alongside with the poorest optical quality of the defective Ge buffer.

By increasing the temperature, the PL features experience a well-known red-shift as a result of the bandgap shrinking, while the concomitant integrated PL intensity discloses a non-trivial temperature behavior that differs drastically for the three $\mathrm{Ge_{0.91}Sn_{0.09}}$ epilayers. Figure 2d  surprisingly unveils that the coherent indirect $\mathrm{Ge_{1-x}Sn_x}$ film demonstrates the simple monotonous PL quenching commonly argued to pertain to direct bandgap materials. Furthermore, the other two partially relaxed samples exhibit a puzzling surge of the PL intensity as the lattice temperature turns higher than 90 K ($\mathrm{T^{-1} \sim 1.1\times 10^{-2} K^{-1}}$). Interestingly, such a steep increase gains a larger dynamic range by increasing the strain relaxation. We emphasize that since the three $\mathrm{Ge_{0.91}Sn_{0.09}}$ epitaxial layers differ solely by the thickness, in other words by the strain, their remarkable distinction in terms of temperature characteristics under the same generation rate stems from the activated optical activity of the dislocations injected in the two thickest samples through the plastic strain relief. 

The results summarized in Figure 2d for the partially relaxed samples compare well with recent reports about the optical properties of Ge/Si epitaxial architectures \cite{Pezzoli14, Pezzoli_arXiv} and prove that in Ge-rich $\mathrm{Ge_{1-x}Sn_x}$ alloys the defect-assisted recombination obeys the Sch\"{o}n-Klasens thermal mechanisms.\cite{Reshchikov14} We can therefore rationalize the physical processes leading to the results of Figure 2 by considering the following scenario. At low temperatures, carriers photogenerated in indirect bandgap $\mathrm{Ge_{1-x}Sn_x}$ alloys can be efficiently trapped at dislocations, if present, because their energy levels lie within the band gap of the host semiconductor. Above a certain critical temperature, however, the thermal activation of trapped carriers occurs, eventually strengthening the radiative interband recombination, as experimentally observed in Figure 2b for the dislocated samples. 

In light of these findings, we point out that the measurement of the temperature-dependent PL characteristics as a function of the epilayer thickness, and specifically the rise of the band-edge recombination, can provide a sensitive spectroscopic method for the determination of the critical thickness above which dislocation nucleation is energetically favorable compared to the elastic energy accumulation in the growing film. This is of central importance for the ultimate fabrication of devices relying on heteroepitaxy on lattice mismatched substrates.

A closer look to the PL data of Figure 2 provides us with additional information about the recombination pathways introduced by the defects in the $\mathrm{Ge_{1-x}Sn_x}$ alloys. The negligible emission intensity at the low-energy side of the main PL peaks discloses a low quantum efficiency for the radiative recombination at dislocations. This makes the experimental investigation of the defect-assisted carrier dynamics more subtle and challenging than in Ge. This elemental semiconductor belongs to the vast class of semiconductors governed by the Sch\"{o}n-Klasens mechanisms,\cite{Reshchikov14, Pezzoli14} where a defect-related PL peak can be often observed and its measurable temperature-dependent quenching is accompanied by a coincident rise of other PL features.\cite{Pezzoli14} It is therefore tempting to conceive that in $\mathrm{Ge_{1-x}Sn_x}$ alloys the suppression of the dislocation-related PL might originate from the peculiar nature of the electronic states therein associated to the extended defects. Indeed our structural data are in line with the available literature reports \cite{Gallagher15, Wirths15} and suggest the prominence of the $\mathrm{90^o}$-dislocations, which are less common than the $\mathrm{60^o}$-dislocations typically distinguished in Ge-based heterostructures.      

It is worth noticing that another consistent explanation  can be put forward for the low emission efficiency of the dislocation-related PL. By alloying Ge with Sn, the offset between the direct and indirect conduction band minima decreases. The more direct-like character of the band structure of $\mathrm{Ge_{1-x}Sn_x}$ compared to the one of Ge is expected to shorten the radiative lifetime. Under this condition, the interband transitions become more probable, and consequently the radiative recombination through the band edges can increase at the expenses of the carrier recombination events, including radiative ones, taking place at the dislocation sites. 

All these arguments call for dedicated and systematic investigations, presently beyond the scope of this work, aimed at the experimental determination of the minority lifetimes in the binary alloy system and at a better theoretical understanding of the so-far elusive electronic states of defects in group IV semiconductors.

In the following, we further substantiate our assertion that the measurement of the temperature-dependent PL, and thereby the recombination dynamics, is strongly affected by the presence of optically-active dislocations. To this purpose, we consider the set of indirect bandgap $\mathrm{Ge_{1-x}Sn_x}$ heterostructures having the lowest Sn content, namely 5\% (see Table 1). 

\begin{figure}
\includegraphics[width=8.5cm]{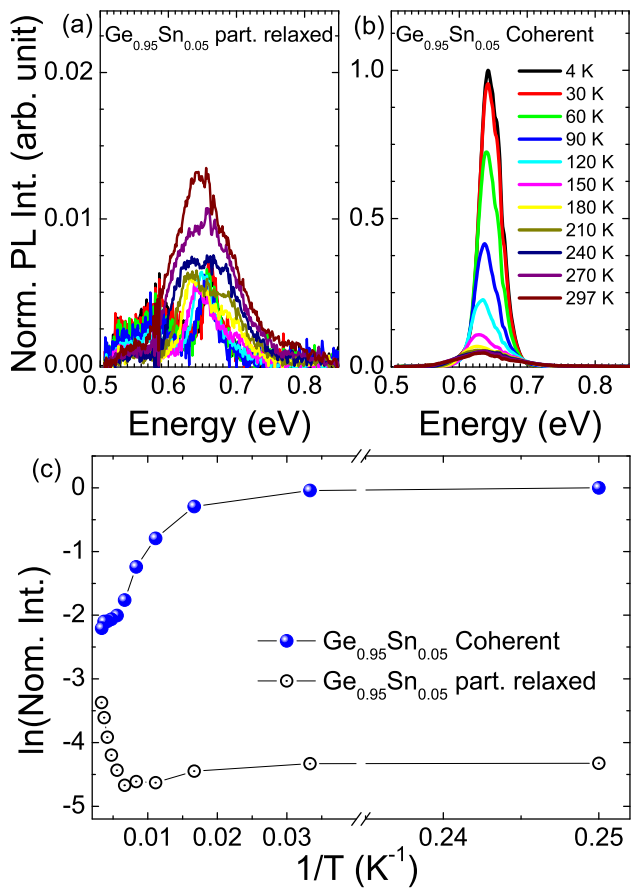}
 \caption{Photoluminescence (PL) spectra as a function of the temperature of (a) a partially relaxed and (b)  a fully strained coherent $\mathrm{Ge_{0.95}Sn_{0.05}}$ layers epitaxially grown on Ge buffered Si(001) substrates. (c) Arrhenius plot summarizing the temperature dependence of the integrated PL intensity for the partially relaxed (open dots) and coherent (full dots) films. All the data are normalized to the low temperature emission intensity of the coherent $\mathrm{Ge_{0.95}Sn_{0.05}}$ sample.}
	\label{fig3}
 \end{figure}

A direct comparison between Figure 2 and Figure 3 elucidates that the $\mathrm{Ge_{0.95}Sn_{0.05}}$ epilayers fully recover the optical properties previously discussed for the Sn-richer $\mathrm{Ge_{0.91}Sn_{0.09}}$ heterostructures. In particular, the temperature characteristic again demonstrates a net monotonic PL quenching as the thickness of the topmost $\mathrm{Ge_{1-x}Sn_x}$ alloy layer is reduced so that coherent growth is satisfied and plastic relaxation inhibited. 

Figure 3c harbors final intriguing phenomena. The deep in the integrated PL intensity curve of the partially relaxed $\mathrm{Ge_{0.95}Sn_{0.05}}$ epilayer occurs at a temperature of about 120 K ($\mathrm{T^{-1} \sim 8\times 10^{-3} K^{-1}}$), thus higher than what obtained for the dislocated $\mathrm{Ge_{0.91}Sn_{0.09}}$ samples. The increase of such critical temperature with decreasing the Sn content reflects the concomitant increase of the bandgap amplitude: A finding that naturally arises in the framework of rate equations modeling the capture and thermal release of carriers by the charged defect lines.\cite{Pezzoli14, Figielski78}

In conclusion, by studying the temperature-dependent PL of indirect bandgap $\mathrm{Ge_{1-x}Sn_x}$ epitaxial layers in the vicinity of the plastic strain-relief, we were able to gather direct insights into the optical activity of growth defects, namely dislocations. The unveiled pivotal role of dislocations on the recombination dynamics suggests that steady-state PL measurements need to be complemented by the direct measurements of the carrier lifetime in order to precisely resolve the directness of the electronic band structure. Our experimental findings can guide and stimulate novel theoretical investigations of the electronic structures of dislocations in Ge-based heterostructures, eventually providing a deeper understanding of existing and future experiments. Finally, in view of practical applications, we anticipated an advanced contact-less spectroscopic technique to pinpoint dislocation injection in lattice mismatched heteroepitaxy.\\


The author thanks E. Vitiello and S. De Cesari for technical assistance, M. Guzzi and E. Grilli for fruitful discussions. This work was supported
by the Fondazione Cariplo through Grant No. 2013-0623.


%


\end{document}